\documentclass[11pt]{cernrep}
\usepackage{graphicx}
\usepackage{here}
\usepackage{amsmath}
\newcommand{\Vek}[1]{\mbox{\boldmath${#1}$}}
\newcommand{\dr}{\text{d}}      
\newcommand{\dd}{\,\text{d}}    
\begin{document}
 \title{AN UNFOLDING METHOD FOR HIGH ENERGY PHYSICS EXPERIMENTS}
\author{Volker Blobel}
\institute{Institut f\"ur Experimentalphysik,
Universit\"at Hamburg, Germany}
\maketitle
\begin{abstract}
Finite detector resolution and limited acceptance require to apply
unfolding methods in high energy physics experiments. Information on the
detector reso\-lution is usually given by a set of Monte Carlo events.
Based on the  experience with a widely used unfolding program (RUN) 
a modified method has been developed.

The first step of the method is a maximum likelihood fit of the
Monte Carlo distributions to the measured distribution in one, two or
three dimensions; the finite statistic of the Monte Carlo  events is taken
into account by the use of Barlows method with a new method of solution.
A clustering method is used before to combine bins in sparsely populated
areas.
In the second step a regu\-larization is applied to the solution,
which introduces only a small bias. The regularization parameter is
determined from the data after a diagonalization and rotation
procedure.
\end{abstract}

\section{THE UNFOLDING PROBLEM} \label{sec:unfolding}

A standard task in high energy physics experiments is the measurement
of a distribution $f(x)$ of some kinematical quantity $x$. 
With an ideal detector one could measure the
quantity $x$ in every event and could obtain $f(x)$ by a
simple  histogram  of  the  quantity   $x$. 
With \emph{real}  detectors  the  determination  of  $f(x)$  is
complicated by three effects:

\begin{itemize}
\item  {\bf Limited acceptance:}
The probability to  observe  a  given  event,  the 
\emph{detector acceptance}, is less than 1. The acceptance 
depends on the kinematical variable $x$. 
\item  {\bf Transformation:}
Instead of the quantity $x$  a  different,  but  related
quantity $y$ is measured.
The transformation from $x$ to $y$  can be  caused  by  the
non-linear response of a detector component.
\item  {\bf Finite  resolution:}
The measured quantity  $y$ is  smeared out  due  to
the finite resolution (or limited measurement  accuracy)
of the  detector.  Thus  there  is  only  a {\em statistical}
relation between the true kinematical variable  $x$  and
the measured quantity $y$.
\end{itemize}

The really difficult effect in the data correction for
experimental effects, or data transformation from $y$ to $x$
is the {\bf finite resolution},
causing a \emph{smearing} of the
measured quantities.
Mathe\-ma\-tically  the
relation between the distribution  $f(x)$  of  the  true
variable $x$, to be determined in an experiment, and the
measured distribution $g({y})$ of  the 
measured  quantity $y$   is  given  by  the
integral equation,
\begin{equation}   \label{eq:basiceq} \boxed{
  g({y})  =
 \int     A({y},x) f(x)  \dd x }   \; ,
\end{equation}
called a Fredholm integral equation of the  first  kind.
In practice often a known  (measured or simulated)
 background contribution $b({y})$ has to
be added to the right-hand side of equation \eqref{eq:basiceq}; 
this contribution is ignored in this paper.
 The  resolution  function  $A({y},x)$
represents the effect  of  the  detector.  For  a
given value $x = x_0$ the function  $A({y},x_0)$
describes the response of the detector  in  the
variable ${y}$ for that fixed  value  $x_0$.  The
problem to  determine  the  distribution
$f(x)$ from measured  distributions  $g({y})$ is called \emph{unfolding};
it is called an inverse problem. Unfolding  of
course requires the knowledge of the resolution function
$A({y},x)$,  i.e.  all  the  effects   of   limited
acceptance, transformation and finite resolution.

In addition to the imperfections  of  the  detector,
there  may  be further
effects between  $x$  and  ${y}$,  which  are 
\emph{outside of the experimental control}, even with an  ideal
detector. One example are radiative  effects,  which  in 
experiments  are  often   corrected   afterwards 
\emph{(radiative  corrections)},  but  are  in their effect
similar to detector effects. If  the
true kinematical quantity is defined at the
\emph{parton level}, further effects from the  fragmentation  process
of partons to the  (observable)  hadrons  influence  the
measured quantity ${y}$. All these  effects  are  of
statistical nature. 

\begin{figure}[!htb] \begin{center}
\includegraphics[width=8cm]{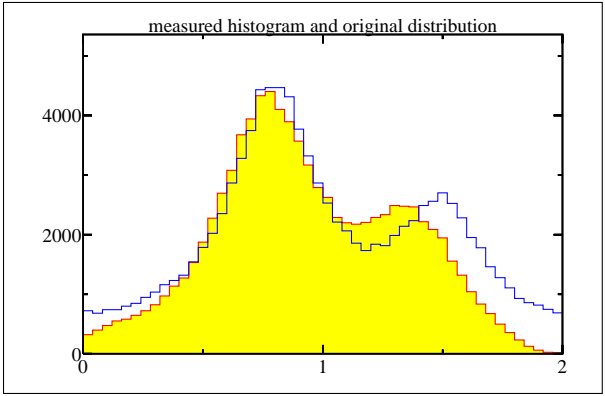}\hspace{0.5cm}\begin{minipage}[b]{7.5cm}
\caption{The Monte Carlo simulation of the effects of limited
acceptance, transformation and finite resolution. Shown is the 
original (true) distribution (line histogram)
 and the ''measured'' distribution
(yellow/shaded histogram).
\label{fig:fmeas}}
\end{minipage}
\end{center}
\end{figure}

A typical example for these effects is shown in Figure \ref{fig:fmeas}
taken from a Monte Carlo simulation of all three effects. By unfolding
an estimate of the original distribution has to be determined
from the distorted measured distribution. 
Details on the Monte Carlo simulation are given later in section
\ref{sec:proposed}, where the unfolding in this example is discussed
in detail.

For the numerical solution of equation \eqref{eq:basiceq} the distributions
have to represented by a finite set of parameters. One posssibility
is to represent the distributions by histograms, and the resolution
function by a matrix. Equation \eqref{eq:basiceq} can then be represented
 by the matrix equation
\begin{equation} \label{eq:basicvec}
            \boxed{
            \Vek{y} =  \Vek{A} \Vek{x}  }
            \; ,
\end{equation}
which has to be solved for the vector $\Vek{x}$, given the vector
$\Vek{y}$ (data histogram).
The vector $\Vek{y}$ with $n$ elements represents a
histogram of the measured quantity $y$, and the distribution $f(x)$ 
is represented by a histogram of the vector $\Vek{x}$ with $m$ elements.
The variables $y$ and $x$ may be multidimensional, and the multidimensional
histograms can be mapped to $n$-bin ($\Vek{x}$) 
 and $m$-bin histograms ($\Vek{y}$), 
respectively. 
The transition from $\Vek{x}$ to $\Vek{y}$ is described by the $n$-by-$m$
matrix $\Vek{A}$.  
The element $a_{ij}$ is related to the probability to observe  an  entry  in
histogram bin $i$ of the histogram $\Vek{y}$, if the  true
value $x$ is from histogram bin  $j$  of  the  histogram
$\Vek{x}$. Problems with standard solutions are discussed in the next
section.

In high energy physics experiments the problems is even more difficult
than in other fields. Often the statistics of the measurement
is not high and every $\Vek{y}$-bin content (which is distributed due to the 
Poisson distribution around the expected value) has a large 
statistical fluctuation. Furthermore the resolution function
$\Vek{A}(x,y)$ (or the matrix $\Vek{A}$) is not known
analytically, 
but is represented by a data set from 
 Monte Carlo simulation of the process, based
on some assumed distribution $f_{\text{MC}}(x)$,
\begin{equation} \label{eq:simul}
  g_{\text{MC}}({y})  = \int     A({y},x) f_{\text{MC}}(x) \dd x   \; ,
\quad \quad \quad \quad \quad \quad \text{(Monte Carlo simulation)}
\end{equation}
and is also statistically limited. Standard methods for the solution of 
integral equations or linear equations can not be used
in this case.  

A simple method like the so-called \emph{bin-by-bin correction}
may be meaningful if the measurements $y$ are very
close to the true values $x$.
Real \emph{unfolding} methods, taking all the correlations into account,
are essential if there are larger effects of \emph{transformation}
and  \emph{finite resolution}. A solution $\Vek{x}$ has to be found, with 
small deviations between the elements of $ \Vek{A} \Vek{x}$ and 
the elements of the actually measured histogram $\widehat{\Vek{y}}$.
 In the maximum likelihood method
a function $F(\Vek{x})$ is constructed as the negative 
log of the Likelihood function, which describes the statistical
relations between data and result:
\begin{equation}   \label{eq:maxlik}
                 F(\Vek{x}) = - \log L(\Vek{x},\Vek{y},\Vek{A})
\end{equation} 
and the minimum of $F(\Vek{x})$ is determined.  
Wildly fluctuating results $\Vek{x}$
are due to large (negative) correlations between adjacent bins
 and  are not acceptable.
The approach to get a more reasonable solution
is to impose a measure of the smoothness on 
the  result $\Vek{x}$; this method is called {\bf regularization}.
This technique was proposed independently by Phillips  \cite{phil} and
by Thikhonov \cite{tik63}.
For a function $f(x)$ the integrated square of the second 
derivative
\begin{equation}   \label{eq:regfun}
          C(f) = \int \left( \frac{\dr^2 f}{\dr x^2} \right)^2 
                 \dd x
\end{equation}
is often used in the regularization 
which in the linearized version of the problem can be expressed
by a quadratic form 
$C(\Vek{x}) = \Vek{x}^T \Vek{C} \Vek{x}$
with a positive-semidefinite matrix $\Vek{C}$ (derivatives are replaced
by finite differences). Equation \eqref{eq:maxlik} 
is then modified to the form
 \begin{equation}   \label{eq:maxlikreg}
                 F(\Vek{x}) = - \log L(\Vek{x},\Vek{y},\Vek{A})
                              + \tau \cdot  C(\Vek{x})
\end{equation}
with a factor $\tau$ called regularization parameter.

The result of the minimization of the modified function
$ F(\Vek{x})$ of equation \eqref{eq:maxlikreg} will show smaller fluctuations
than the result obtained from equation \eqref{eq:maxlik}
and may be more useful
to compare the measurement with theoretical predictions.    
However it is clear that unavoidably the regularization introduces
a bias. The magnitude of the bias depends on the value of regularization
parameter $\tau$.  A very large value would result 
in a \emph{linear} function  $f(x)$ or distribution $\Vek{x}$, 
respectively. It is clear that the method requires an
a-priori knowledge about a smooth behaviour of $f(x)$. The 
function $f_{\text{MC}}(x)$ used in the Monte Carlo simulation
of equation \eqref{eq:simul} is often very close to the final result $f(x)$, 
i.e. the ratio is rather smooth. This suggests to express
$f(x)$ in the form $f(x) = f_{\text{MC}}(x) \times f_{\text{mult}}(x)$
and to rewrite equation \eqref{eq:basiceq} in the form
\begin{equation}   \label{eq:basiceqm}
  g({y})  =
 \int  \left[ A({y},x) f_{\text{MC}}(x)  \right]
      f_{\text{mult}}(x)  \dd x  \; .
\end{equation}
For the discretized form 
the function $f_{\text{MC}}(x)$ can be absorbed in a redefinition
of matrix $\Vek{A}$ and the vector $\Vek{x}$ is interpreted as 
discretization of the hopefully \emph{smooth} function $f_{\text{mult}}(x)$.
With this redefinition the equation \eqref{eq:basicvec} can 
remain unchanged.     
The program RUN \cite{blocern,run} for regularized unfolding
is available since almost
two decades and has been used in many experiments; early applications are
\cite{jonker}  and \cite{berger}. It is based on the 
reinterpretation of matrix $\Vek{A}$ and  $\Vek{x}$, as described above, 
and includes a method for the determination of the regularization
parameter $\tau$ based on the available degrees of freedom. 
In the method described later in this paper some details are treated
differently.

\section{UNFOLDING AS AN ILL-POSED PROBLEM} \label{sec:ill}

The problems inherent to unfolding are discussed in a simple
special case,
assuming a resolution matrix $\Vek{A}$ with some smearing of data into 
neighbour bins. Assuming a true vector $\Vek{x}$ the product
$\Vek{y}= \Vek{A} \Vek{x}$ describes the distribution expected due to
the migration effect. With the same dimensions for the vectors
$\Vek{x}$ and $\Vek{y}$  the matrix  $\Vek{A}$ is a square matrix and 
in the example later in 
 this section the following simple symmetric form is assumed for the
matrix  $\Vek{A}$, which depends on a single parameter
$\varepsilon$ ($\varepsilon$ = migration parameter); for a 
5-by-5 matrix the form is
\begin{equation}  \label{eq:matra}
 \Vek{A} =  \left(
        \begin{array}{ccccc}
 1-\varepsilon  & { \varepsilon}      & 0 & 0 & 0 \\
  \varepsilon   & { 1-2  \varepsilon} &  \varepsilon & 0 & 0 \\
    0        & { \varepsilon} & 1-2  \varepsilon &  \varepsilon & 0 \\
  0   & 0  & \varepsilon & 1-2  \varepsilon & \varepsilon \\
0 & 0 & 0 & \varepsilon & 1-\varepsilon 
\end{array}     \right)  \; .
\end{equation}
A direct solution for $\Vek{x}$, given a measurement 
$ \widehat{\Vek{y}}$  differing  from the expectation
$\Vek{A} \Vek{x}$ with the true vector $\Vek{x}$ by statistical
fluctuations,  is possible with inversion of the matrix $\Vek{A}$:
\begin{equation*}
\text{estimate} \quad \widehat{\Vek{x}} = \Vek{A}^{-1} \widehat{\Vek{y}}
\quad \quad  \quad \quad \quad \text{error propagation} \quad
\Vek{V}(\widehat{\Vek{x}}) = \Vek{A}^{-1} \Vek{V}_y
                      \left(  \Vek{A}^{-1} \right)^T  \; .
\end{equation*}
The result has certain good statistical properties,
 for example it has no bias:
$ E \left[ \Vek{\widehat{x}} \right] = \Vek{A}^{-1} E\left[ \Vek{y} \right]
  = \Vek{A}^{-1}  \Vek{A}  E \left[ \Vek{x} \right]  
  =  \Vek{x}$. 
In practice the result is however satisfactory only for a matrix $\Vek{A}$
with dominating diagonal; the result looks terrible if the matrix
$\Vek{A}$ describes a large migration to neighbour bins.
Consequently the problem is called an ill-posed problem.
In the following the solution of the equation $\Vek{y} =  \Vek{A} \Vek{x}$
using an orthogonal decomposition is discussed; this will allow some insight 
into the unfolding problem.

\noindent
The symmetric matrix $\Vek{A}$ is expressed by
\begin{equation}  \label{eq:udut}                           
      \Vek{A}  =  \Vek{U}  \Vek{D} \,  \Vek{U}^T
\end{equation}
with a transformation matrix $\Vek{U}$
with property $\Vek{U}^T  \Vek{U} = \Vek{1}$, and a diagonal matrix $\Vek{D}$, 
where the diagonal elements of  matrix $\Vek{D}$ are the 
eigenvalues $ \lambda_j$ of matrix $\Vek{A}$  
(in the order of decreasing value).
The transformation matrix $\Vek{U}$ contains the corresponding eigenvectors
with  the eigenvector $\Vek{u}_j$ in the  $j$-th column. 
The condition number $\kappa$ of a matrix is defined by the ratio
of eigenvectors  $\kappa = \lambda_{\text{max}}/\lambda_{\text{min}}$;
the value of $\kappa$  is important for the quality of unfolding (see
below).
For values above  $\varepsilon = 0.20$ the condition
number $\kappa$ is very rapidly increasing.  

A transformation of equation $\Vek{y} =  \Vek{A} \Vek{x}$
 to a new basis is done
by multiplication with matrix $\Vek{U}^T$ (which is 
a rotation in the $n$-dimensional space): 
\begin{eqnarray*}
\Vek{U}^T \cdot \;\;| \qquad \qquad \qquad
  \Vek{y} &=& \Vek{A} \Vek{x} = \Vek{U} \Vek{D} \Vek{U}^T \Vek{x} \; . \\
\underbrace{{ \Vek{U}^T  \Vek{y}}}_{=\Vek{c}} &=& 
  \Vek{D} \underbrace{ { \Vek{U}^T \Vek{x}}}_{=\Vek{b}} 
\quad \quad \quad \quad \quad \to \quad \quad \Vek{c} = \Vek{D} \Vek{b} \; .
\end{eqnarray*}
The matrix $\Vek{U}$ of eigenvectors $\Vek{u}^T_j$ allows to
transform the vectors $\Vek{x}$ and $\Vek{y}$ to vectors
 $\Vek{b} = \Vek{U}^T\Vek{x}$ and $\Vek{c} = \Vek{U}^T\Vek{y}$, 
and to transform these vectors back by 
$\Vek{x} = \Vek{U}\Vek{b}$  and $\Vek{y} = \Vek{U}\Vek{c}$. 
The transformed equation $\Vek{c} = \Vek{D} \Vek{b}$ with
the diagonal matrix $\Vek{D}$ shows, that each of the 
coefficients $b_j$ and $c_j$ is transformed independently
of any other coefficient by the simple relation
$c_j = \lambda_j \cdot b_j$. This operation does not depend on any
assumption of the solution $\Vek{x}$, and depends only on 
the properties of the matrix $\Vek{A}$.
 Folding ($\Vek{x} \to \Vek{y}$) and unfolding 
($\Vek{y} \to \Vek{x}$) is multiplication and division by the 
eigenvalues $\lambda_j$, respectively, of the coefficients in the
transformed space.

In order to unfold a measured vector  $\Vek{y}$, the vector is
transformed by $\Vek{c} = \Vek{U}^T \Vek{y}$ to coefficients $c_j$,
which have values influenced by statistical fluctuations of the elements
of vector $\Vek{y}$. In the unfolding the coefficients $c_j$ are divided
by the eigenvalues $\lambda_j$ to obtain $b_j = c_j/\lambda_j$; 
 the statistical fluctuation of 
coefficient $c_j$ is magnified for small eigenvalues $\lambda_j$ 
(i.e. $\lambda_j \ll 1$).
Eventually, for very small eigenvalues $\lambda_j$, the final result 
$\Vek{x} = \Vek{U} \Vek{b}$ will be dominated by one or by few of the  
coefficients $b_j$ with small eigenvalues and large statistical errors, and 
the complete result is unsatisfactory.

\begin{figure}[!htb] \begin{center}
\includegraphics[width=5.3cm]{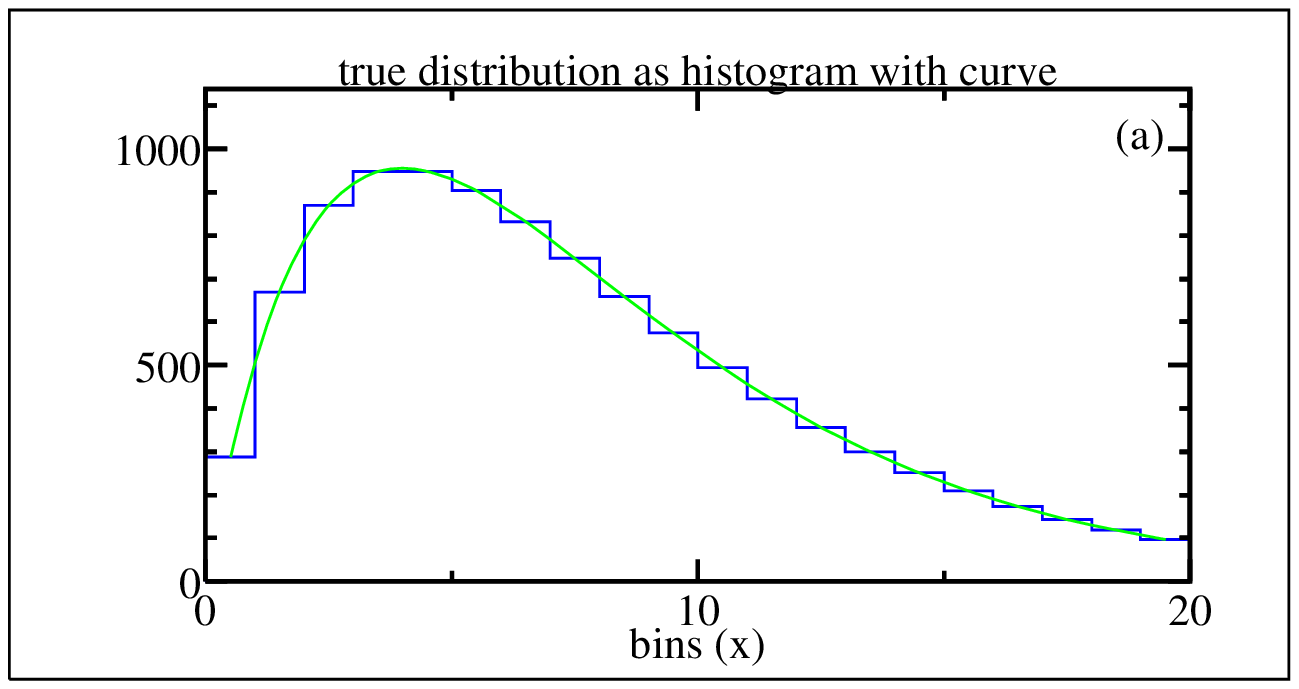}%
\includegraphics[width=5.3cm]{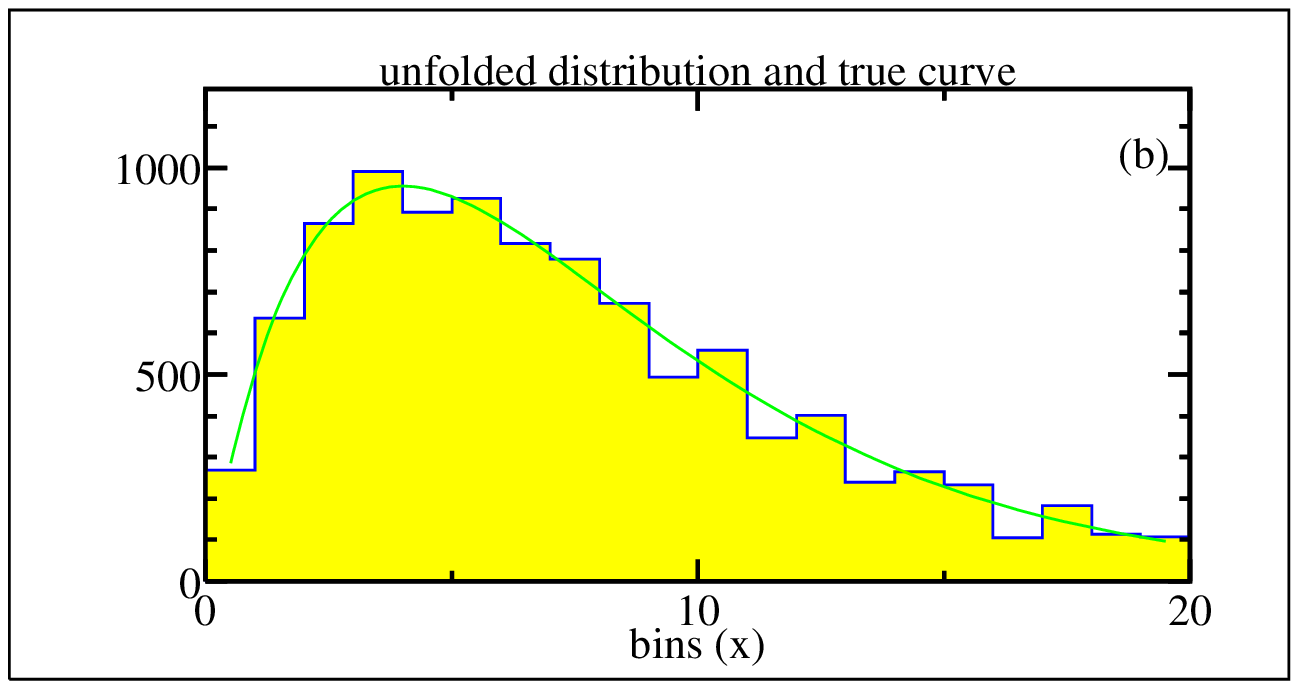}%
\includegraphics[width=5.3cm]{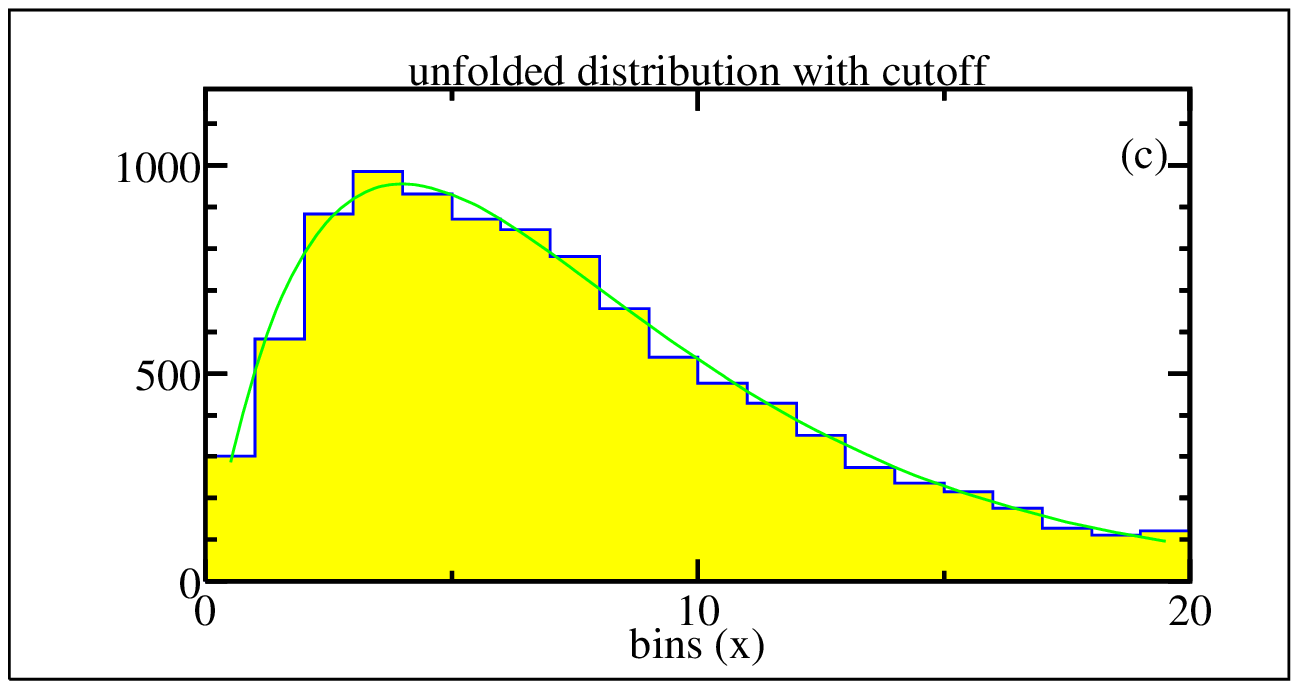}  
\caption{Original (true) distribution (a) and two results from 
unfolding ((b) and (c)). Result (b) has been obtained
from all 20 coefficients, and for result (c) a sharp cut-off after
10 coefficients has been applied (i.e. the coefficients 11 to 20 
are ignored). 
\label{fig:trueunf}}
\end{center}
\end{figure}

\begin{figure}[!tb] \begin{center}
\includegraphics[width=8cm]{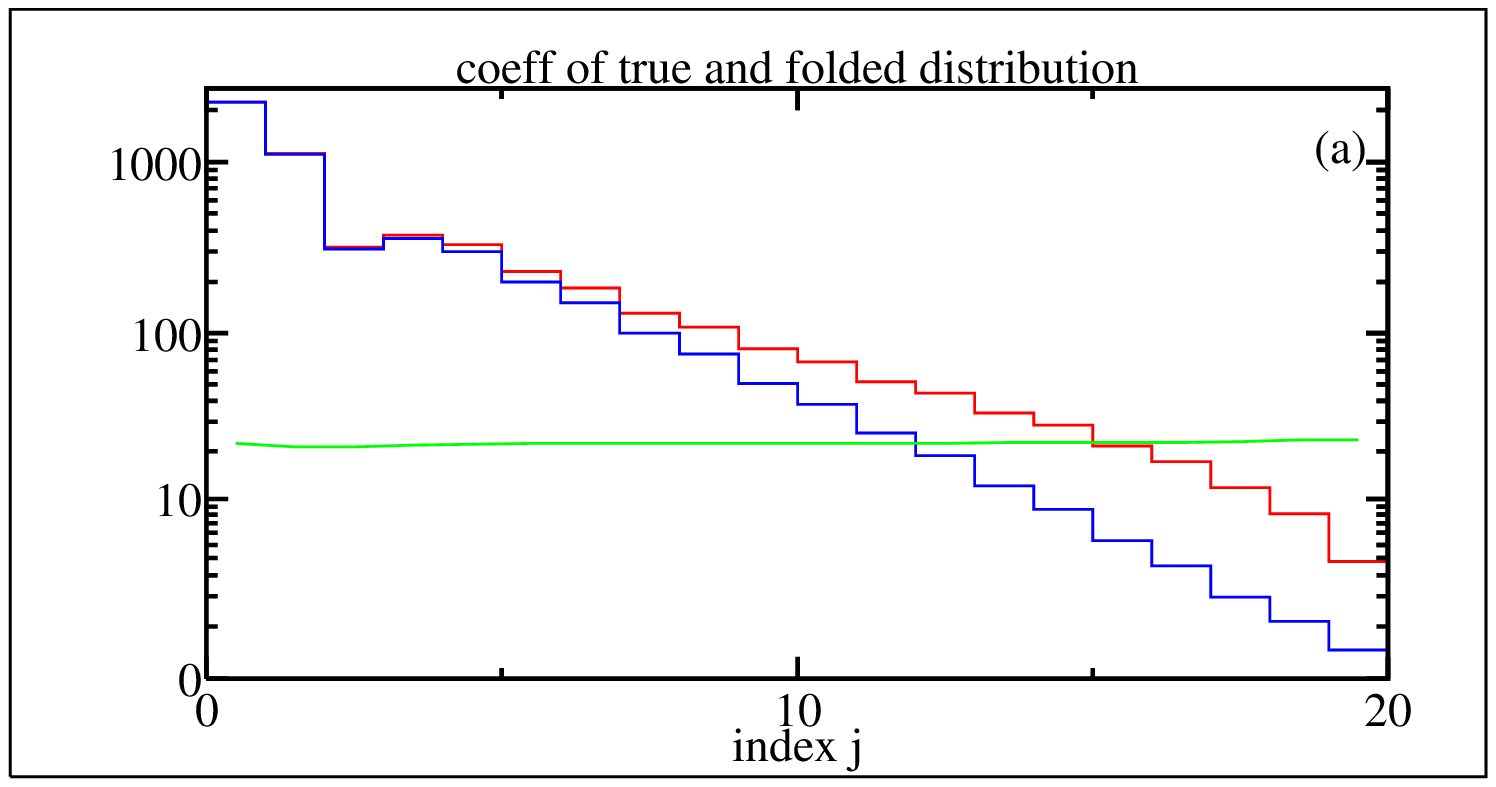}%
\includegraphics[width=8cm]{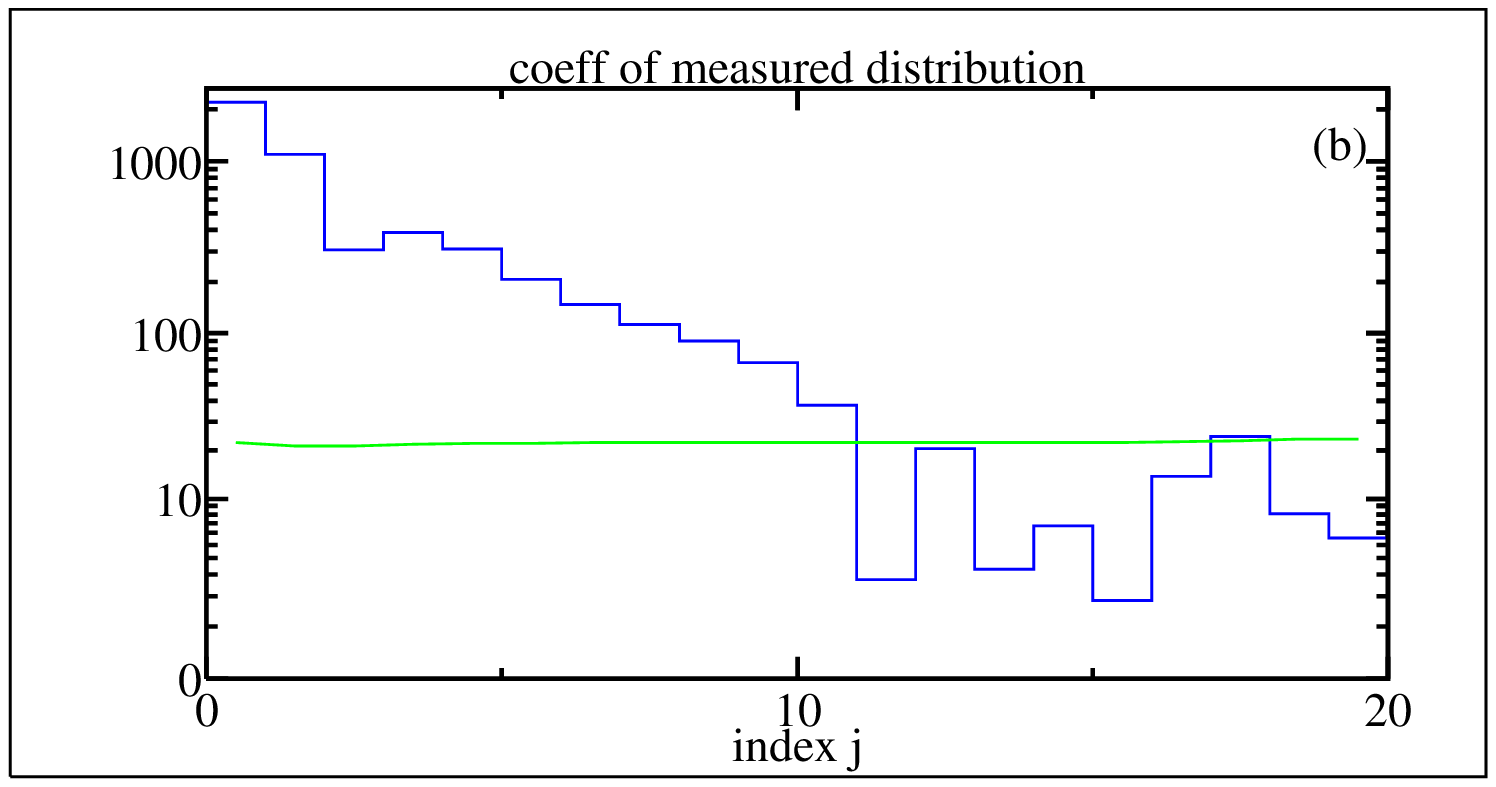}  
\caption{The absolute values of 
 coefficients $b_j$ and $c_j$ are shown for $j=1,\, 2,
\ldots 20$. The coeffients  $b_j$ and $c_j$ for the true distribution 
and the folded distribution (without measurement errors) are shown 
in (a), together with the (almost constant)
 error estimate for the coeffients $b_j$
calulated by error propagation. The coeffients  $c_j$ from 
the simulated measured distribution are
shown in (b), together with the error estimate. 
For  $j$ above $12$ the smaller coefficients of the folded distribution
 become smaller than the statistical error. In (b)
the coefficients for  $j$ above $12$ are dominated by statistical 
errors and even the sign is not determined by the data.
\label{fig:coeffs}}
\end{center}
\end{figure}

\noindent
{\bf Example.} In a numerical example the matrix $\Vek{A}$ has the 
form of equation \eqref{eq:matra} with $n=20$ and a value of the migration
parameter of $\varepsilon=0.22$. The first eigenvalue is $\lambda_1 =1.0$,
and the last one is $\lambda_{20} = 1/7.9$, giving a condition number
$\kappa = 7.9$.
For $\Vek{x}$ the ideal distribution of
Figure \ref{fig:trueunf}a is assumed; the underlying function is of the 
form $x \exp(-a x)$. 
The decomposition of the matrix  $\Vek{A}$ according to equation
\eqref{eq:udut} is performed and the coefficients $b_j$ and $c_j$ are
calculated.
These coefficients are shown in Figure \ref{fig:coeffs}a
(with $b_j \ge c_j$). In addition
this figure shows, calculated by standard error propagation, the 
almost constant 
error level of the coefficients, of the folded distribution of 
Figure \ref{fig:trueunf}a with Poisson distributed bin contents. 
Figure \ref{fig:coeffs}a  shows,
that the coefficients $b_j$ of the true distribution decrease 
rapidly with increasing value $j$ of the index of the coefficient,
by roughly three orders of magnitude. The coefficients $c_j$ of the folded
distribution drop even faster, because it is more smooth due to 
the migration effect.  
Of course the relation $b_j/c_j = \lambda_j$ is valid.  
The last coefficient $b_j$ in Figure \ref{fig:coeffs}a is reduced
to $c_j$
by the inverse of the condition number of the matrix, which is 
$\kappa = 7.9$ in this case.

The components
of the first eigenvector $u_1$ (eigenvalue = 1) are all the same.
Thus the coefficients $b_i$ and $c_i$ are identical, and proportional
to the total sum of the measured distribution, not at all influenced by the
migration. If visualized by functions, interpolating the
components the eigenvector $u_j$
(eigenvalue $\lambda_j$) has $j-1$ zeros, and the curvature
of the visualized eigenvectors is rapidly increasing with index $j$.
The components of the last eigenvector $u_n$  have alternating sign
for the bins; it
has a small
absolute value and its measured value
 will have a large relative statistical error.
The value of $b_{20}$ is obtained by $b_{20} = 7.9 \cdot c_{20}$ in  
 unfolding,
introducing a large bin-to-bin oscillation into the result of
unfolding.

In a simulation Poisson distributed bin contents are assumed in 
the measurement vector $\Vek{y}$. 
The coefficents for this measured distribution are 
shown in Figure \ref{fig:coeffs}b, together with the level of the
statistical error. As expected from the size of the errors 
all coefficients with an index above about $j=12$ are dominated by the
statistical error and therefore do not significantly contribute to 
the information content of the measurement. 
For indices above  $j=12$ even the sign of the 
coeffient can not be determined by the measurement.   

Using all the ''measured'' coefficients  for the 
unfolding the result of Figure \ref{fig:trueunf}b is obtained.
This 
result shows large fluctuations around the expected values shown by the 
curve. The fluctuations are due to the contributions from indices
above $j=12$, which represent noise and are magnified in the 
unfolding because of the large values of their inverse eigenvalues.
The result is clearly unsatisfactory. 

Because all measured coefficients $c_j$ with $j$ above a value
of 12 are dominated by statistical errors (noise) their use
in the unfolding makes no sense. A sharp cut-off after index $j=12$ or
even after index $j=10$ will not remove any useful information from 
the measurement. The unfolding result using only measured coefficients
$c_j$ up to $j=10$ is shown in Figure \ref{fig:trueunf}c;
compared to Figure \ref{fig:trueunf}b the large fluctuations are suppressed
and the results seems to be acceptable. Of course the fine structure
of the true distribution expressed by the true coefficients
 $b_j$ with $j>10$
is not included in the solution and this may represent a bias. It is 
an unavoidable bias because these coefficients can not be measured. 

The covariance matrix of the result can be calculated by standard
error propagation. 
However it is clear that the covariance matrix is singular and has 
only rank 10 in this case, because the 20 bins are obtained from 
10 measured coefficients (10 degrees of freedom).
This property is inherent to the  cut-off method and to the
regularization method, and was already mentioned in \cite{blocern}.
Such singularity of the covariance matrix can be avoided if  
the final transformation is to a number of bins identical to the 
degree of freedoms; only a limited number of bins can be obtained
in a measurement with large miration effects. 

This method of using a sharp cut-off has to be compared to the
regularization method. It has been shown \cite{blocern}
that the use of a
regularization function of the type of equation \eqref{eq:regfun}
is equivalent to a \emph{smooth} cut-off; essentially the measured 
coefficients $c_j$ are multiplied by a factor depending on the 
curvature of the orthogonal contributions (see section 
\ref{sec:proposed}).\footnote{
Sometimes the iterative solution of unfolding problems expressed by
the equation $\Vek{y} =  \Vek{A} \Vek{x}$ 
 is proposed in the literature without explicit
regularization, starting from a ''good'' initial distribution for 
$\Vek{x}$.
Of course equations of this type (with a square matrix)
have a unique solution and iterative solutions are slow compared to 
the direct solution; after a large number of iterations with convergence
the same unsatisfactory 
 result as by direct solution will be obtained. However
in these proposals only a small number of iterations is recommended.
It can be shown that  iterative methods can in fact include
an implicit regularization \cite{louis}: there is a different 
speed of convergence for the various orthogonal contributions and 
the small contributions with a small eigenvalue will converge very
slowly. Thus after a few iterations the (large) coefficients with large 
eigenvalues are already accurate; the remaining coefficients
are still almost unchanged and thus, for a stop after few iterations, 
their values are still close to the starting values. 
There is of course some subjectivity in stopping ''early'' enough.}

\section{THE PROPOSED UNFOLDING METHOD} \label{sec:proposed}

The proposed method is similar to the method used in RUN;
the differences are emphasized in this section. It is expected that 
the proposed modifications results in more stable solutions. 
The proposed method requires large dimension parameters in the 
resolution matrix $\Vek{A}$. Like in RUN the regularization is determined
by the required number of degrees of freedom, which determines the 
regularization parameter.

Figures in this  section  refer to the example already mentioned
in section \ref{sec:unfolding}
In a Monte Carlo calculation of all three effects, limited
($x$-dependent) acceptance, non-linear transformation and 
finite resolution are simulated. Details on the function and 
the distorting effects are identical to the published examples
\cite{blocern}. In total 100~000 ''events''
are simulated for ''data'' and for the  MC  defining matrix $\Vek{A}$. 
The input function $f_{\text{MC}}(x)$ (equation \eqref{eq:basiceqm})
is a constant.

In RUN the discretization for $f(x)$ and for $A(y,x)$ was done using
cubic B-spline functions;
 the effect is the same as for simple 
histograms namely the integral equation is transformed to a 
system of linear equations, however the elements of the 
vectors are B-spline coefficients instead of bin contents. The  
advantage is that the parametrized solution is a \emph{smooth} function
and the curvature as defined by equation \eqref{eq:regfun} can 
be exactly written as a quadratic form. 
However the accurate determination of matrix $\Vek{A}$ 
requires a good  Monte Carlo statistic. In RUN statistical fluctuations
of the elements of matrix $\Vek{A}$ could not be treated.  

Simple histograms are instead proposed here; the elements of the vector
$\Vek{y}$ are bin contents (integer numbers).
The curvature of the solution is constructed by
finite differences: the second derivative in bin 
$j$ is proportional to $x_{j-1} - 2 x_{j} + x_{j+1}$.
In a histogram  some  resolution is lost if bins with a width as large as
expected for the final resolution would be used. It is 
recommended to use initially $m=2 n_{\text{df}}$ bins for $\Vek{x}$ for 
a final number of degrees of freedom of $n_{\text{df}}$. For $y$ a larger
number of bins $n$ ($ > m$) is recommended, to avoid a loss of resolution.
Thus the number
$n \times m$ of elements is large, and a large sample of Monte Carlo
events is required to \emph{fill} matrix $\Vek{A}$. The statistical 
error of the elements  $a_{ij}$  eventually can not be neglected.

\noindent
{\bf Standard Poisson maximum likelihood fit.}
Ignoring initially eventual statistical errors of the elements $a_{ij}$
the expected number of events in bin $i$ of $\Vek{y}$ is given by
$y_i = \sum_{j=1}^m a_{ij} \, x_j$ .
For the expected number $y_i$, as given by this expression,
the observed values $\widehat{y}_i$ follows the Poisson distribution.  
Optimal estimates for the elements $x_j$ are obtained by 
minimizing the (negative) logarithm of the total
likelihood  with respect to the elements $x_j$ of vector $\Vek{x}$, 
assuming the Poisson distribution: 
\begin{equation}   \label{eq:fmaxlik}
     F(\Vek{x}) = - \ln {\cal L}(\Vek{x}) = 
    - \ln  \left[   
       \prod_{i=1}^n  P_{y_i}(\widehat{y}_{i})
           \right]   
  = \sum_{i=1}^n \left(
 y_i - \widehat{y}_i \cdot \ln y_i     \right)   + \text{const.} \; ,
\end{equation}
where the constant term containing e.g $\widehat{y}_i!$ can be ommited. 
This expression \eqref{eq:fmaxlik} 
 correctly accounts also for bins with a small number of
histogram entries 
$\widehat{y}_i$. 

An alternative would be to use the (linear) least squares method 
with singular value decomposition
for the fit.  However for small number of entries
the use of the Poisson distribution seems to be essential. Furthermore 
the diagonalization used later in the method is almost 
equivalent to singular value decomposition (eigenvalues are the 
squares of the singular values).

\noindent
{\bf Fitting with finite Monte Carlo samples.}
The problem of statistical fluctuations of the elements 
$a_{ij}$ has been neglected so far.  A method to treat 
the problem within the maximum-likelihood method has been developed
by R.Barlow and Chr.Beeston \cite{barlow}. 
In this method there is for each source bin $x_j$ 
some (unknown) expected number of events $A_{ij}$. 
For each element $A_{ij}$ the corresponding number
$a_{ij}$ from the Monte Carlo sample is generated by a
distribution which is taken to be Poisson too. The nice feature of this 
method is that a bias which would be introduced by ignoring the
statistical character of the values of the elements $a_{ij}$ is avoided
and the maximum likelihood error is more realistic. A large 
number of slack variables (one for each bin) is introduced and 
has to be treated in the optimzation.  
A new fast numerical solution method has been
developed (see \cite{home}).

\noindent
{\bf Combining bins.}
The likelihood function is a sum over all bins.
Combining almost empty bins does not introduce a
systematic error.  The total number of elements of the matrix
 may be large,
especially if $x$ and/or $y$  are multidimensional, and a small number of
entries (or even zero) in an element may not be uncommon.
The combination of almost empty bins is done with a cluster algorithm,
taking into 
account the distance between bins in one, two or three dimensions.

\noindent
{\bf First option: sharp cut-off of orthogonal contributions.}
This method is rather similar to the method discussed in section 
\ref{sec:ill}. The computational problem is to determine the 
minimum of $F(\Vek{x})$ (see equation \eqref{eq:fmaxlik}). 
The standard iterative method is based on the representation
for the correction $\Delta \Vek{x}$
\begin{equation}   \label{eq:fdeltax}
     F(\Delta \Vek{x}) = \frac{1}{2}  \Delta \Vek{x}^T \Vek{H} 
   \Delta \Vek{x} 
  +  \Delta \Vek{x}^T \Vek{g} + \ldots
\end{equation}
with the Hessian $\Vek{H}$ (matrix of second derivatives of
$F(\Delta \Vek{x})$) and the gradient vector \Vek{g} (first 
derivatives of $F(\Delta \Vek{x})$). A Newton step is then calulated
from equation
$ \Vek{H}   \Delta \Vek{x}  +  \Vek{g}  =0$. 
Convergence is usually fast for good starting values and the 
covariance matrix is equal to the inverse of the Hessian.
The starting 
values can be calculated by a linear least square fit, based on the 
approximation of the Poisson distribution by a Gaussian distribution
for each bin.

A sharp cut-off as discussed in the example of section \ref{sec:ill} requires
a diagonalization of the symmetric matrix 
 $\Vek{H}$ by 
$      \Vek{H}  =  \Vek{U}  \Vek{D} \,  \Vek{U}^T$
with a diagonal matrix $\Vek{D}$ and a transformation matrix $\Vek{U}$.
By a transformation (rotation) in $\Vek{x}$-space  linear combinations
of the $\Vek{x}$-components are obtained with a dia\-gonal 
covariance matrix, with variances of the linear combinations given 
by the inverse of the eigenvalues of matrix $\Vek{D}$. A cut-off is done 
at some index $j$ 
followed by backtransformation to the $\Vek{x}$-space of bin-contents 
using the transformation matrix $\Vek{U}$.

\noindent
{\bf Second option: regularization.} 
In this option the regularization is based on the second derivative of the 
result according to equation \eqref{eq:regfun}, which can be expressed
by a quadratic form  $\Vek{x}^T \Vek{C} \Vek{x}$ 
 with a positive-semidefinite matrix $\Vek{C}$. 
In principle the same procedure is used as in RUN; the mathematical 
details are given elsewhere \cite{blocern}.
Here a simple
explanation is given on the standard mathematical
operations\footnote{In a
publication the method has been described to 
''\emph{have certain mathematical complications}'', but it is based
only on
standard linear algebra of symmetric matrices.} used. 
Regularization is done by adding the term
$\tau \cdot \Vek{x}^T \Vek{C} \Vek{x}$ to the function  $F(\Delta \Vek{x})$
of equation \eqref{eq:fdeltax}. Exactly as in the first option
the Hessian is diagonalized. 
\begin{equation}  \label{eq:udutu}
      \Vek{H}  =  \Vek{U}  \Vek{D} \,  \Vek{U}^T  
   \qquad \qquad \qquad
       \Vek{H}^{-1}  =  \Vek{U}  \Vek{D}^{-1}   \Vek{U}^T 
               = \left[ \Vek{U} \Vek{D}^{-1/2} \right] 
                 \left[ \Vek{D}^{-1/2} \Vek{U}^T \right]  \; .
\end{equation}   
 Up to this step everything is 
identical to the cut-off option.
Using transformation matrix $ \Vek{U} \Vek{D}^{-1/2}$ the vector
$\Vek{x}$ is transformed to linear combinations $\widehat{\Vek{x}}$, 
which are orthogonal, with all variances equal to 1 (unit covariance
matrix).
Because the covariance matrix
is equal to the unit matrix, every additional pure rotation
will not change the (unit) covariance matrix. In terms of the  
transformed vector the regularization term can now be written in the
form $\tau \cdot \widehat{\Vek{x}}^T C_U \widehat{\Vek{x}}$, where 
$C_U$ is the transformed curvature matrix $\Vek{C}$. Now another
diagonalization can be done of matrix $C_U$: 
\begin{equation}
\tau \cdot  \Vek{x}^T \Vek{C} \Vek{x} \;  \to  \;
\tau \cdot \widehat{\Vek{x}}^T \Vek{C}_U \widehat{\Vek{x}} 
= \tau \cdot \widehat{\Vek{x}}^T  \Vek{U}_C \; \Vek{S} \; 
            \Vek{U}_C^T \widehat{\Vek{x}}  
\end{equation} 
with a diagonal matrix $\Vek{S}$ and a rotation matrix $\Vek{U}_C$.
This diagonalization can be used to define a pure rotation from the 
linear combination $\widehat{\Vek{x}}$ to another linear combination
$\widetilde{\Vek{x}}$
\begin{equation}
   \widehat{\Vek{x}} \;  \to  \;
    \widetilde{\Vek{x}} =   \Vek{U}_C^T \, \widehat{\Vek{x}} \; .
\end{equation}
The components of the new vector $\widetilde{\Vek{x}}$ still have
the unit matrix as covariance matrix. The complete transformation
from $\Vek{x}$ to $\widetilde{\Vek{x}}$ is the effect of the 
transformation by  $ \Vek{U} \Vek{D}^{-1/2}$   and by $\Vek{U}_C$.
The algebra can be explained 
in other words: the error ellipsoid related to the Hessian is first
rotated to have the axes parallel to the axes of the new system. By 
a change of the scales the ellipsoid is transformed to a sphere, 
which will remain a sphere for any further rotation. 
A last rotation is done to bring the axes into the order of 
increasing curvature.  

\begin{figure}[!h] \begin{center}
\includegraphics[width=8cm]{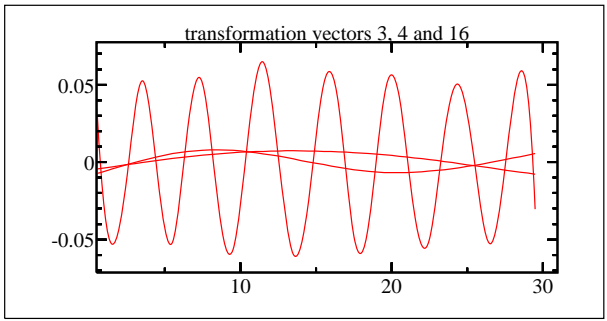}\hspace{0.5cm}\begin{minipage}[b]{7.5cm}
\caption{Selected column vectors of the complete transformation matrix
defined in the regularization procedure. They correspond to the 
curvature eigenvalues $S_{33}$, $S_{44}$ and $S_{16,16}$. 
Visualization is done by curves  interpolating the components. 
The amplitude associated which each vector all have
the same standard deviation of 1. 
\label{fig:feig}}
\end{minipage}
\end{center}
\end{figure}

\begin{figure}[!htb] \begin{center}
\includegraphics[width=8cm]{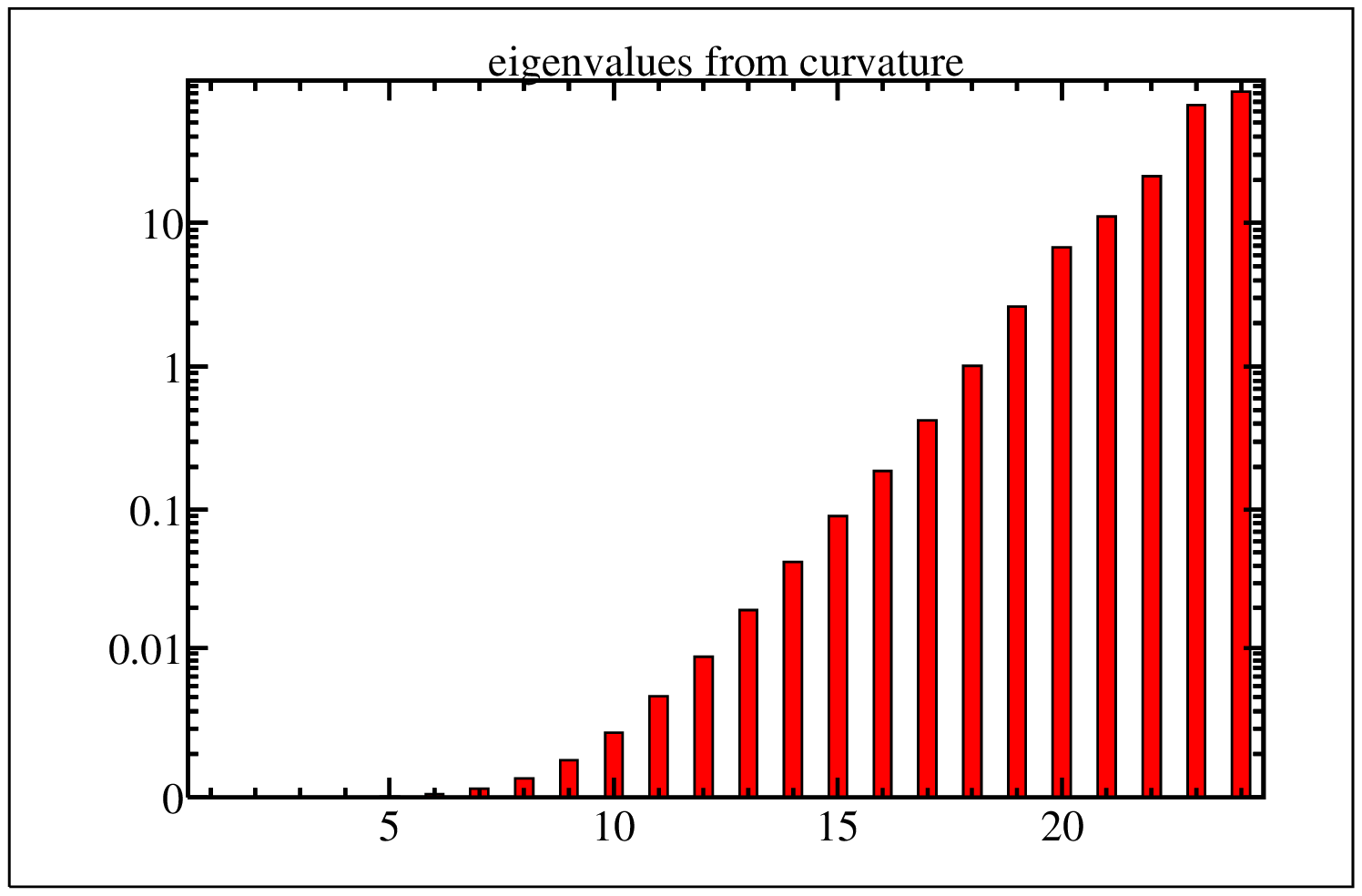}%
\includegraphics[width=8cm]{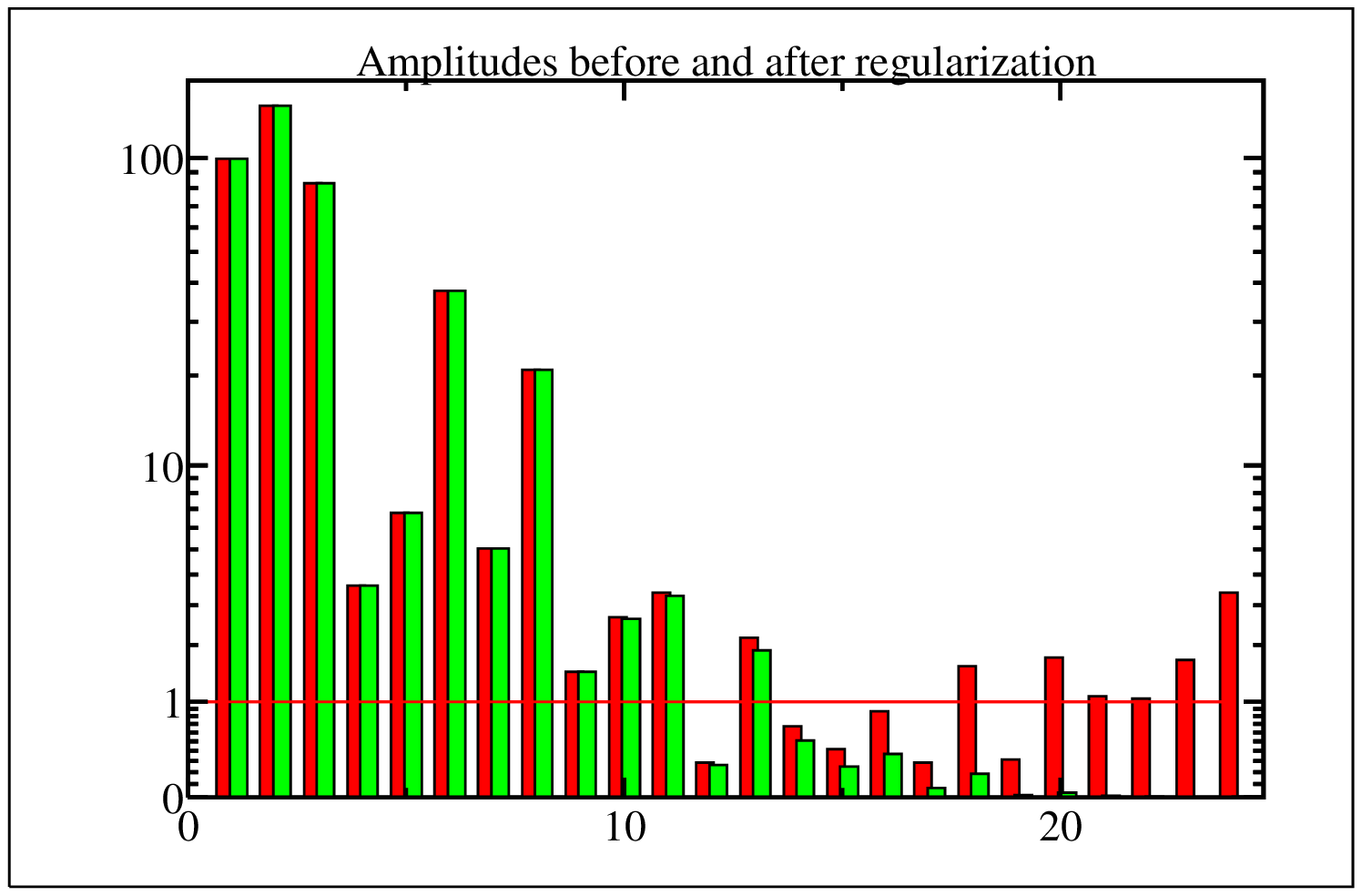}  
\caption{The eigenvalues after the curvature transformation.
The values are very rapidly increasing for orthogonal contributions
for increasing index value (left).
The amplitudes before (left bars) and after regularization 
(right bars). The statistical error of all amplitudes is equal 
to 1, which is indicated by the horizontal line.
The vertical scale is linear at the bottom and makes a transition 
to a logarithmic scale at the top (right). \label{fig:famp}
\label{fig:eicu}}
\end{center}
\end{figure}

Some columns of the complete (product) transformation are shown in Figure
\ref{fig:feig}. All linear combinations obtained have the same
precision (standard deviation of the coefficient is one). As seen
in the Figure linear combinations with large index $j$ are oscillating 
with large amplitude.
 The diagonal elements
$S_{jj}$ are the (statistically independent) contributions of the elements
of  $\widetilde{\Vek{x}}$ to the total curvature. Sorted according
to increasing value of $S_{jj}$ the value of $S_{jj}$  will increase rather
fast with increasing index $j$.
The spectrum of eigenvalues $S_{jj}$ is shown
in Figure \ref{fig:eicu}. 
In terms of the linear combinations  $\widetilde{\Vek{x}}$ 
regularization is simply given by 
\begin{equation} \label{eq:regul} 
      \left(    \widetilde{x}_j \right)_{\text{reg}}
      = \left( \frac{1}{1 + \tau \cdot S_{jj}} \right)
  \left( \widetilde{x}_j
         \right)_{\text{unreg}}   \; .
\end{equation}
and this simple form is the reason for the transformations made before.

\noindent
{\bf Determination of the regularization parameters $\tau$.}
 The first factors (small $j$) on the right-hand-side of equation 
\eqref{eq:regul} will be close
to 1; for a value $\tau = 1/S_{kk}$ the factor will be 1/2 and 
for indices $j > k$ will rapidly decrease towards zero.
The sum of all factors can be called the \emph{effective} number of degrees
of freedom, and can be used to determine the value
of the regularization parameter $\tau$ from the required number of
degrees of freedom, i.e. the regularization parameter
 $\tau$ is determined from the value of $n_{\text{df}}$ in the  equation
\begin{equation}    
   n_{\text{df}} = \sum_{j=1}^m
 \left( \frac{1}{1 + \tau \cdot S_{jj}} \right) \; .
\end{equation}
Thus the required number of degrees of freedom has to be specified and 
determines the degree of regu\-lari\-zation. This
number can be taken from the spectrum  of the 
coefficients or amplitudes, shown in Figure \ref{fig:famp}. 
 The insignificant part (large $j$) is clearly
visible in the spectrum and separated from the significant part 
(small $j$). The selected value of  $n_{df}$ should be equal to or larger
than the number of significant terms.
The unregularized
amplitudes, which have standard deviation one, are shown by the left bars; 
amplitudes above index 15 are compatible with one and represent noise.
They would however make a large contribution to the solution, because
the corresponding column vectors (Figure \ref{fig:feig}) are large.
The regularization effectively damps the amplitude
(right bars)  around and above 
index 15, which has been chosen as the degree of freedom here. 
The significant amplitudes are not affected by the regularization.

\begin{figure}[!h] \begin{center}
\includegraphics[width=8cm]{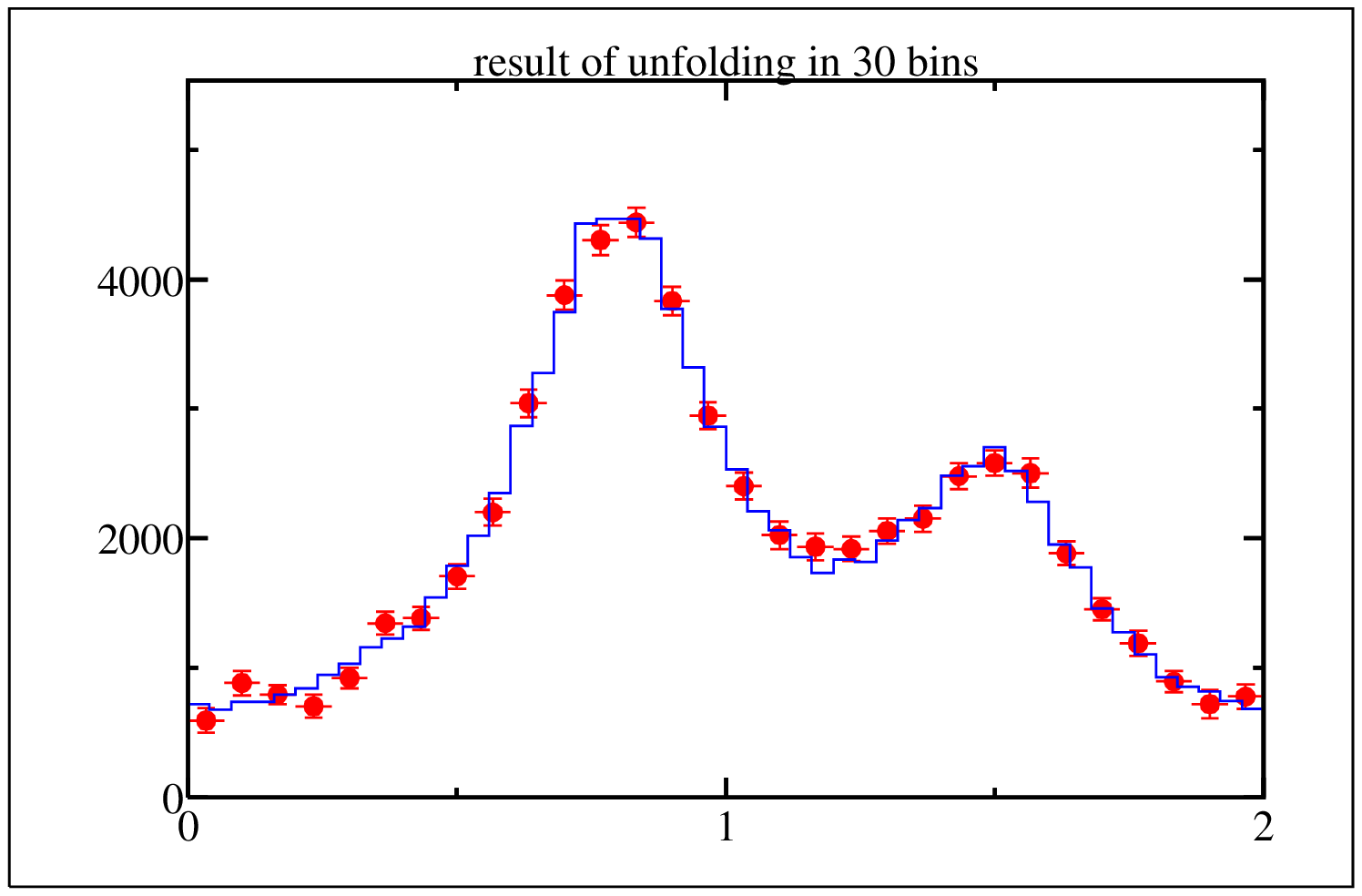}%
\includegraphics[width=8cm]{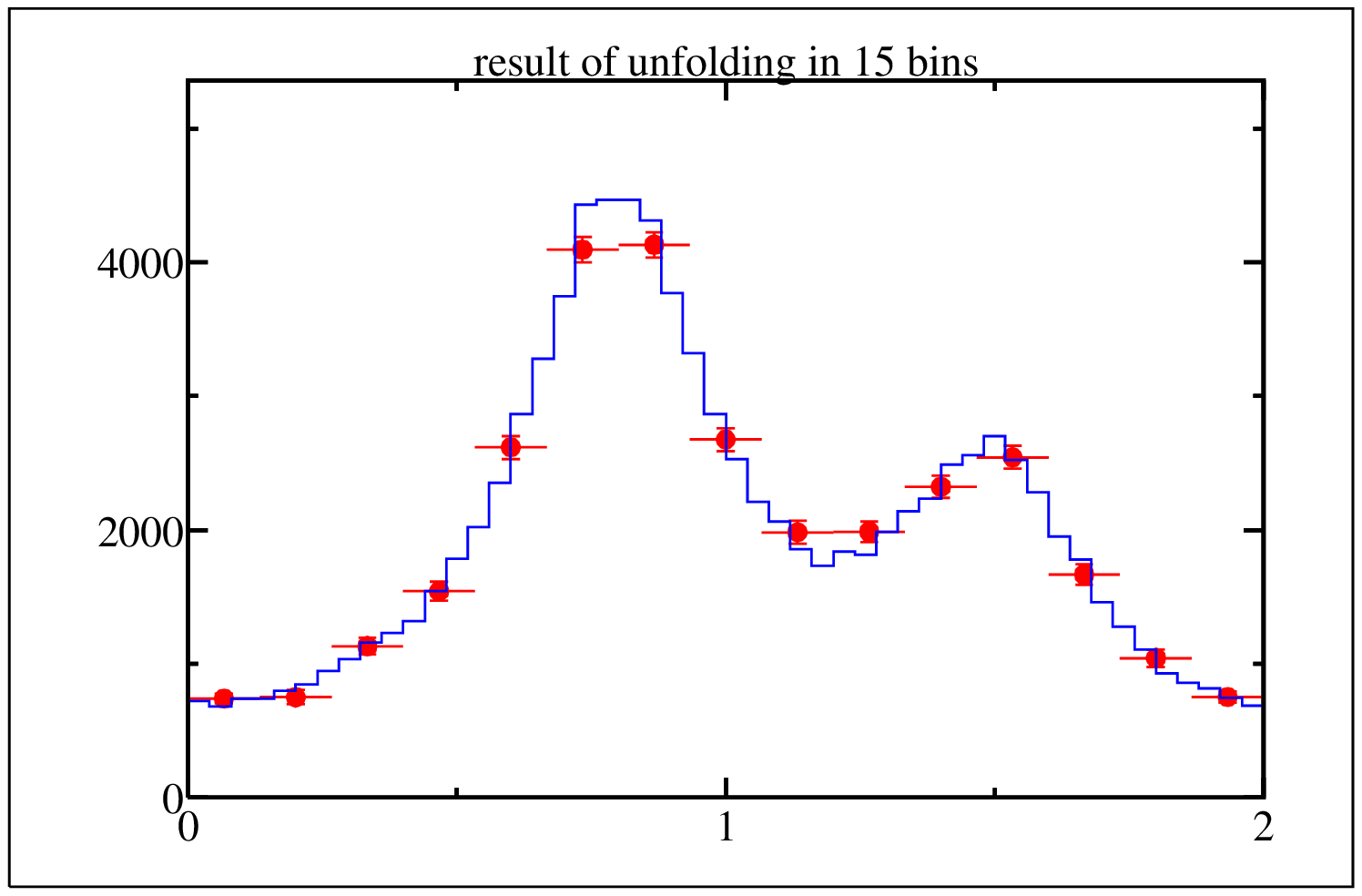}
\caption{The unfolding result after regularization with 15 degrees
of freedom with 30 bins (left) and with 15 bins (right). For comparison
the original (true) distribution is shown by a histogram.
The data from Figure \ref{fig:fmeas} are used as input.
\label{fig:fres}}
\end{center}
\end{figure}

The final result of the example (measured distribution in Figure
\ref{fig:fmeas}) 
 is shown in Figure \ref{fig:fres}.
The left figure shows 30 data points with error bars together with
the original (true) distribution; within errors the original 
distribution is nicely reproduced. The rank of the covariance 
matrix is about 15, which was chosen as the effective number of 
degrees of freedom; thus inversion of the covariance matrix,
needed e.g. for a least-square fit of a model to the data, is not
possible. Although the large number of 30 data points seems to be
attractive, the data points should be reduced to 15 data points by combining
two bins to one, which then have a full-rank covariance matrix.
This set of data points is shown in Figure \ref{fig:fres} (right). 
The broader bins of this set of data points are a consequence of the 
limited acceptance and finite resolution of the measurement.  


\section*{ACKNOWLEDGEMENTS}

I would like to thank the organizers of the conference on 
Advanced Statistical Techniques in Particle Physics
for their hospitality and the stimulating atmosphere in Durham.

\end{document}